\documentclass[twocolumn]{aa}
\usepackage[varg]{txfonts}

\usepackage{graphicx}
\usepackage{natbib}
\usepackage{color}
\usepackage{subfig}
\usepackage[switch]{lineno}
\usepackage{xspace}
\usepackage{microtype}
\usepackage{amssymb}
\usepackage{mathtools}
\usepackage{multirow}
\def\link_col{blue}
\usepackage{url}
\usepackage{wasysym}
\usepackage{caption}

\def\gray{$\gamma$-ray\xspace}
\def\grays{$\gamma$-rays\xspace}
\def\fermi{{\it Fermi}-LAT\xspace}

\def \msun{\mbox{\it M$_{\odot}$}\xspace}

\def\deg{\hbox{$^\circ$}}
\def\arcmin{\hbox{$^\prime$}}

\begin{document}

\title{Diffuse \gray emission toward the massive star-forming region, W40}
\titlerunning{\fermi detection toward W40}
\author{Xiao-Na Sun\inst{1,2}
\and Rui-Zhi Yang\inst{3,4,5}
\and Yun-Feng Liang\inst{6}
\and Fang-Kun Peng\inst{7}
\and Hai-Ming Zhang\inst{1,2}
\and Xiang-Yu Wang\inst{1,2}
\and Felix Aharonian\inst{8,9}
}
\institute{School of Astronomy and Space Science, Nanjing
University, Nanjing 210093, China
\and Key laboratory of Modern Astronomy and Astrophysics, Nanjing
University, Ministry of Education, Nanjing 210093, China
\and Department of Astronomy, School of Physical Sciences, University of Science and Technology of China, Hefei, Anhui 230026, China
\and  CAS Key Labrotory for Research in Galaxies and Cosmology, University of Science and Technology of China, Hefei, Anhui 230026, China 
\and School of Astronomy and Space Science, University of Science and Technology of China, Hefei, Anhui 230026, China 
\and Laboratory for Relativistic Astrophysics, Department of Physics, Guangxi University, Nanning 530004, China
\and Department of Physics, Anhui Normal University, Wuhu, Anhui 241000, China
\and Dublin Institute for Advanced Studies, 31 Fitzwilliam Place, Dublin 2, Ireland
\and Max-Planck-Institut f\"ur Kernphysik, Saupfercheckweg 1, 69117 Heidelberg, Germany
}

\abstract {
We report the detection of high-energy \gray signal towards the young star-forming region, W40. Using 10-year Pass
8 data from the  Fermi Large Area Telescope (\fermi), we extracted an extended \gray excess region with a significance of $\sim 18 \sigma$. The radiation has a spectrum with a photon index of $2.49 \pm 0.01$. The spatial correlation with the ionized gas content favors the hadronic origin of the \gray emission. The total cosmic-ray (CR) proton energy in the \gray production region is estimated to be the order of $10^{47}\ \rm erg$. However, this could be a small fraction of the total energy released in cosmic rays (CRs) by local accelerators, presumably by massive stars, over the lifetime of the system. If so, W40, together with earlier detections of \grays from Cygnus cocoon, Westerlund~1, Westerlund~2, NGC~3603, and 30~Dor~C, supports the hypothesis that young star clusters are effective CR factories. The unique aspect of this result is that the \gray emission is detected, for the first time, from a stellar cluster itself, rather than from the surrounding "cocoons". 
}
\maketitle

\section{Introduction}
For decades, supernova remnants (SNRs) have been believed to be the main contributors to the Galactic CRs.  The detection of \grays from a number of SNRs do confirm their ability to accelerate particles to high and very high energies. At the same time, the recent discoveries of $\gamma$-ray signals from representatives of other source populations, as well as the impressively precise measurements of primary and secondary CRs, indicate the need for modification of the standard paradigm of Galactic CRs. These data demand the existence of additional contributors to the Galactic CRs. In particular, there is growing evidence  that the young star clusters constitute an important class of factories of Galactic CRs  \citep{Aharonian19}. In several such systems, namely the Cygnus cocoon \citep{Ackermann11, Aharonian19}, Westerlund 1 \citep{Abramowski12}, Westerlund 2 \citep{Yang18}, NGC 3603 \citep{Yang17}, and 30 Dor C \citep{Abramowski15}, extended \gray (from GeV to TeV) structures with hard energy spectra have been detected.

\begin{figure}[ht]
\centering
\includegraphics[scale=0.4]{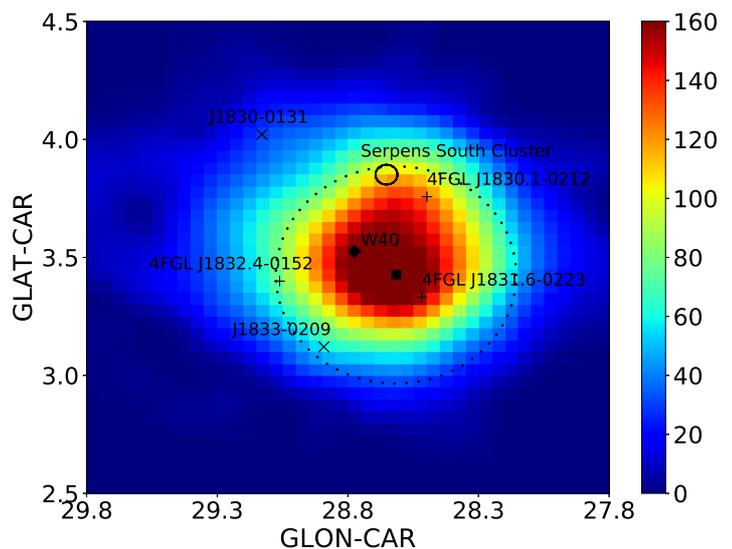}
\caption {TS residual map above 1 GeV in the $2\deg \times 2\deg$ region near W40, with pixel size corresponding to $0.05\deg \times 0.05\deg$, smoothed with a Gaussian filter of $0.45^ \circ$.
The square marks the best-fit central position of the assumed uniform disk used for spatial analysis, and the dashed circle with a radius of $0.46\deg$ shows its size.
The diamond indicates the radio position of W40, and the small circle with a radius of $0.042\deg$ marks the core of the Serpens South Cluster \citep{Gutermuth08}.
The "+" indicate the point sources listed in the 4FGL, and excluded in our background model.
The "$\times$" symbol shows pulsars located close to W40 region in projection on the sky.
}
\label{fig:tsmap}
\end{figure}

There are more than a dozen young star clusters that have not been explored so far in the \gray band \citep[see, e.g., ][]{zwart10,davies12}. One of the reasons is that most of such systems are located inside the Galactic plane, and thus cannot be resolved due to the crowded environments and the limited resolution of \gray telescopes. To this regard, the young star-forming region W40, located three degrees above the Galactic plane, is a unique target for \gray observations.  The W40 complex is one of the nearest sites of massive star formation. The distance to  W40  complex is $436.0 \pm 9.2$ pc derived from a parallax measurement with the VLBA \citep{Ortiz17}.  It is composed of a cold molecular cloud \citep{Goss70} with an area of approximately one square degree. Adjacent to this molecular cloud is a large blister \ion{H}{ii} region \citep{Westerhout58,Vallee87} with a diameter of $\sim 6\arcmin$ and dynamical age of 0.19 - 0.78 Myr \citep{Mallick13}. There is a dense stellar cluster within W40 in the IR maps of the Two Micron All Sky Survey (2MASS) and DENIS surveys \citep{Rey02}. This cluster is dominated by four bright OB stars that are believed to be the primary excitation sources for the nearby \ion{H}{ii} region \citep{Zeilik78,Smith85,Shuping12}.
 Toward the $\sim 20\arcmin$ north-west to the W40 region in the Galactic coordinate, there is another star formation region.
The young Serpens South cluster was found close to the center of this region \citep{Gutermuth08} and is marked in \figurename~\ref{fig:tsmap}.
This cluster consists of a large fraction of protostars, some of which blow out collimated molecular outflows \citep{Nakamura11}.
\cite{Nakamura14, Nakamura17,Shimoikura18} suggested that the two regions might be physically connected based on molecular mapping observations. Initially, the star formation region was found in the W40 region, then the young Serpens South cluster was triggered to form in interaction with the expanding shell of the W40 \ion{H}{ii} region \citep{Shimoikura18}.
The velocity field around the two regions has a complex composition containing several distinct velocity components; the local standard of the rest velocity $V_{\rm LSR}$ is roughly in the range of 0 - 15 km/s \citep{Shimoikura18}.

In this paper, we perform a detailed analysis based on the 10-year \fermi data toward W40. The paper is organized as follows. In Sect.~\ref{sec:fermi_analy}, we present the details of the data analysis. In Sect.~\ref{sec:Gas}, we describe the gas distribution in the vicinity of W40. In Sect.~\ref{sec:cr}, we discuss the possible radiation mechanisms of the \gray emission. In Sect.~\ref{sec:pro}, the CR content around this region is discussed. In Sect.~\ref{sec:conc}, we present the main implications of the obtained results.
%
%


%
%



\section{\fermi data analysis}
\label{sec:fermi_analy}
We selected the \fermi Pass 8 database towards the W40 region from August 4, 2008 (MET 239557417) until April 30, 2019 (MET 578338401) for the analysis.
We used both the front and back converted photons.
A $10\deg \times 10\deg$ square region centered at the position of W40 (R.A. = 277.86\deg, Dec. = -2.07\deg) is chosen as the region of interest (ROI).
We used the "source" event class, recommended for individual source analysis, and the recommended expression $\rm (DATA\_QUAL > 0) \&\& (LAT\_CONFIG == 1)$ to exclude time periods when some spacecraft event affected the data quality.
To reduce the background contamination from the Earth's albedo, only the events with zenith angles under $100\deg$ were included in the analysis.
We processed the data through the current Fermitools from conda distribution\footnote{https://github.com/fermi-lat/Fermitools-conda/} together with the latest version of the instrument response functions (IRFs) P8R3\_SOURCE\_V2.
We used the python module that implements a maximum likelihood optimization technique for a standard binned analysis\footnote{\url{https://fermi.gsfc.nasa.gov/ssc/data/analysis/scitools/python_tutorial.html}}.

In our background model, we included the sources in the \fermi eight-year catalog \citep[4FGL,][]{Fermi19} within the region of ROI enlarged by $7\deg$.
We left the normalizations and spectral indices free for all sources within $6\deg$  distances from W40.
For the diffuse background components, we used the latest Galactic diffuse model gll\_iem\_v07.fits and isotropic emission model iso\_P8R3\_SOURCE\_V2\_v1.txt\footnote{\url{https://fermi.gsfc.nasa.gov/ssc/data/access/lat/BackgroundModels.html}} keeping their normalization parameters free.

\subsection{Spatial analysis}
\label{sec:spatial_analy}
For the study of the spatial distribution of  \gray emission, we used only the events with energy above 1 GeV. We note there are three 4FGL catalog sources (4FGL J1830.1-0212, 4FGL J1831.6-0223, and 4FGL J1832.4-0152) on top of the W40 complex as marked in \figurename~\ref{fig:tsmap}. To study the excess \gray emission around W40, we excluded these three 4FGL sources from our background model. 
We used the gttsmap tool to evaluate a $2\deg \times 2\deg$ residual TS map by removing the contribution from the all known sources in our background model defined above. The TS value for each pixel is set as ${\rm TS}=-2({\rm ln}L_0-{\rm ln}L_1)$, where $L_0$ is the maximum-likelihood value for the null hypothesis, and $L_1$ is the maximum likelihood with an additional source located in this pixel.
As shown in \figurename~\ref{fig:tsmap}, a strong \gray excess near the W40 position is apparent after the fitting and subtraction of \grays from the background sources by performing the likelihood analysis.
We added a point-like source model encompassing the W40's position into our background model, and optimized the localization using the gtfindsrc tool.
The derived best-fit position of the excess above 1 GeV is [R.A. = 277.84\deg, Dec. = -2.24\deg] (the "square" in \figurename~\ref{fig:tsmap}), with 2$\sigma$ error radii of 0.1\deg, and $0.19^\circ$ away from the radio position of W40 \citep{Westerhout58,Sharpless59}. 
Under the assumption of a single point-like source model (Model 1), the significance of the excess \gray emission is TS = 214 ($\sim 15\sigma$).

We considered several uniform disk templates centered at the above best-fit position with various radii from $0.3\deg$ to $0.6\deg$ in steps of $0.02^\circ$ to investigate the spatial extension of the \gray emission. We compared the overall maximum likelihood of the uniform disk ($L$) (alternative hypothesis) with that of the single point-like source model ($L_{0}$) (null hypothesis), and defined the significance of the disk model $-2({\rm ln}L_{0}-{\rm ln}L)$ following the paper by \citet{Lande12}. The alternative hypothesis is significantly preferred to the null hypothesis only if $-2({\rm ln}L_{0}-{\rm ln}L) > 16$. The spectral types of all the added sources for the likelihood ratio test are assumed to be a simple power law.
We found that the uniform disk template with a radius of $\rm R_{disk} \sim 0.46\deg \pm 0.02\deg$ (Model 2) can  best fit the \gray excess, and the improved significance is $\rm \Delta TS=103\ (\sim 10.2\sigma)$ with one additional degree of freedom (dof) relative to the single point-like source model.
The photon index is $2.49 \pm 0.01$, and the total \gray flux is estimated as $(4.43 \pm 0.27) \times 10^{-9} \ \rm ph.cm^-2.s^{-1}$ above 1 GeV.  The associated uncertainties correspond  to the 68\% statistical errors.  Considering the distance of about $430$ pc, the total \gray luminosity is estimated to be $(6.4 \pm 0.4) \times 10^{32} \rm erg/s$. 

To study whether the extended nature of the GeV emission is caused by a superposition of several separate point-like sources, we removed the uniform disk model and recovered the three deleted 4FGL point-like sources (Model 3).
The maximum-likelihood value listed in \tablename~\ref{tab1} appeared to be smaller than the uniform disk case, even with more free parameters. Thus, the three point source scenario is disfavored compared to the uniform disk model. 
Both the star formation regions of W40 and Serpens South Cluster lie within the $\sim 0.46\deg$ uniform disk template of the best-fit position, and we cannot yet rule out the possibility that the GeV \gray emissions, at least in part, originate from the Serpens South Cluster. 
Thus, we put  a single point-like source at the position of  the Serpens South Cluster instead of W40 (Model 4), and the most possible combination of two point-like sources (Model 5), namely, one point-like source at the best fit location, and another one at the position of the Serpens South Cluster.
The fit maximum-likelihood values are smaller than  those of Model 2. Thus, the GeV emission originating from the Serpens South Cluster is disfavored compared to the W40 region.
\tablename~\ref{tab1} lists the models and the log (likelihood) values.

\begin{table*}
\centering
\caption{Fitting results for the different models in Sect.~\ref{sec:spatial_analy}.}
\begin{tabular}{lcccccr}
\hline
\hline
        Model &TS & -log(Likelihood) & dof\tablefootmark{a}\\
 \hline
        Model 1 (single point source)& 214& 742513 & 0 & \\ 
\\
        Model 2 ($0.46\deg$ uniform disk) & 317& 742462 & 1 & \\
\\
        Model 3 (three point sources) & 197& 742477 & 4 & \\
\\
        Model 4 (single point source) & 134& 742551 & 0 & \\
\\
        Model 5 (two point sources) & 182& 742494 & 2 & \\

\hline
\hline
\end{tabular}
\tablefoot{
\tablefoottext{a}{Additional degrees of freedom compared to Model 1.}\\
See Sect.~\ref{sec:spatial_analy} for details.
}
\label{tab1}
\end{table*}

\subsection{Spectral and variability analyses}
\label{sec:spectral_analy}
To study the spectral properties of \gray emission towards W40 and the energy distribution of the parent particles,  we fixed the $0.46\deg$ uniform circle disk as the spatial model of the extended emission, and assumed a power-law spectral shape.

\begin{figure}[ht]
\centering
\includegraphics[scale=0.4]{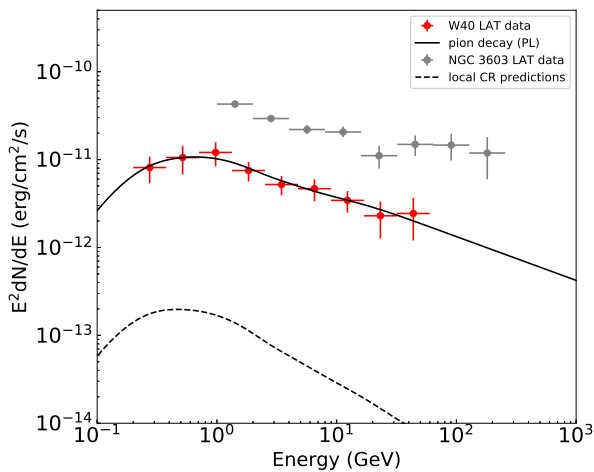}
\caption {
SED of the \gray emission toward W40 for a spatially uniform disk model with a radius of $0.46^{\circ}$.
Both the statistical and systematic errors for the six low-energy bins are considered.
The solid curve represents the spectrum of \grays from interactions of relativistic protons with the ambient gas, assuming a  power-law distribution of protons (see Sect.~\ref{sec:cr}). 
The dashed curve represents the predicted fluxes of \gray emission derived from the \ion{H}{ii} column density map, the CRs are assumed to have the same spectra as measured in the solar neighborhood \citep{Aguilar15}. (For details, see the context in Sect.~\ref{sec:Gas}). The gray data points are the fluxes of NGC 3603 taken from \citet{Yang17}.
}
\label{fig:sed}
\end{figure}

We divided the energy range 200 MeV - 60 GeV into nine logarithmically spaced energy bins and derived the spectral energy distribution (SED) via the maximum-likelihood method. The results are shown in Fig.\ref{fig:sed}.
The significance of the signal detection for each energy bin exceeds $2\sigma$.
We calculated 68\% statistical errors for the energy flux densities.
We also estimated the uncertainties caused by the imperfection of the Galactic diffuse background model by changing the normalization artificially by $\pm$6\% from the best-fit value for each energy bin, and we considered the maximum flux deviation of the source as the systematic error \citep{Abdo09}. 
 
\begin{figure}[ht]
\centering
\includegraphics[scale=0.4]{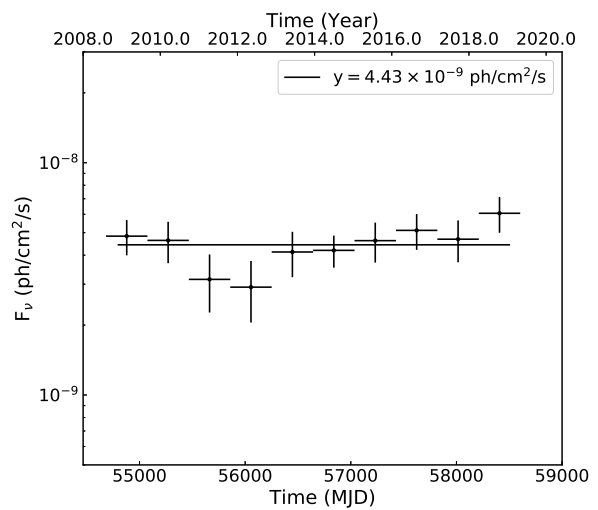}
\caption {
Light curve of \gray emission towards  W40 from August 4, 2008 (MJD 54683) until April 30, 2019 (MJD 58604).
}
\label{fig:lc}
\end{figure}

We tested the possible variability of this source by producing the light curve. This was done by binning the whole data set used into ten equidistant time bins and deriving the \gray spectrum above 1 GeV in each of these bins. The results are shown in Fig.\ref{fig:lc}. We find statistically significant signals in all time bins, without any indication of flux variability. By fitting the light curve with a horizontal line, the derived reduced chi-squared $\rm \chi^2/dof = 0.88$, which is consistent with a constant flux.   
\section{Gas content around W40}
\label{sec:Gas}
We investigated three different gas phases within the  $0.46\deg$ uniform disk shown in \figurename~\ref{fig:tsmap} (hereafter called \gray emission region): the neutral atomic hydrogen (\ion{H}{i}), the molecular hydrogen (H$_{2}$), and the ionized hydrogen (\ion{H}{ii}). We calculated the \ion{H}{i} column densities $N_{\ion{H}{i}}$ from the data-cube of the \ion{H}{i} 4$\pi$ survey (HI4PI), which is a 21-cm all-sky database of Galactic \ion{H}{i} \citep{HI4PI16}.
We used the equation
\begin{equation}
N_{\ion{H}{i}}=-1.83 \times 10^{18}T_{\rm s}\int \mathrm{d}v\ {\rm ln} \left(1-\frac{T_{\rm B}}{T_{\rm s}-T_{\rm bg}}\right),
\end{equation}
where $T_{\rm bg} \approx 2.66\ \rm K$ was the brightness temperature of the cosmic microwave background (CMB) radiation at 21-cm,  and $T_{\rm B}$ is the measured brightness temperature in 21-cm surveys.
In the case of $T_{\rm B} > T_{\rm s} - 5\ \rm K$, we truncated $T_{\rm B}$ to $T_{\rm s} - 5\ \rm K$. A uniform spin temperature $T_{\rm s} = 150 \ \rm K$ was adopted.

We used the carbon monoxide (CO) composite survey \citep{Dame01} to trace the H$_{2}$. The standard assumption of a linear relationship between the velocity-integrated brightness temperature of CO 2.6-mm line, $W_{\rm CO}$, and the column density of molecular hydrogen, $N(\rm H_{2}$), meaning $N({\rm H_{2}}) = X_{\rm CO} \times W_{\rm CO}$ \citep{Lebrun83} were used. The conversion factor $X_{\rm CO}$ was chosen to be $ \rm 2.0 \times 10^{20}\ cm^{-2}\ K^{-1}\ km^{-1}\ s$ \citep{Bolatto13,Dame01}.
We integrated the velocity interval from 0 - 15 km/s based on the results of Fig. $2$ of \citet{Shimoikura18} for the column density calculations of both \ion{H}{i} and H$_{2}$.

The W40 complex also harbors a bright ionized hydrogen region. We adopted the free-free emission map derived from the joint analysis of Planck, WMAP, and 408 MHz observations \citep{Planck16} to obtain the map of \ion{H}{ii} column density.
We first converted the emission measure (EM) into free-free intensity ($I_{\nu}$) by using the conversion factor at 353-GHz in Table 1 of \citet{Finkbeiner03}.
Then we used Equation (5) in \citet{Sodroski97},
\begin{equation}
N_{\ion{H}{ii}} = 1.2 \times 10^{15}\ {\rm cm^{-2}} \left(\frac{T_{\rm e}}{1\ \rm K}\right)^{0.35} \left(\frac{\nu}{1\ \rm GHz}\right)^{0.1}\left(\frac{n_{\rm e}}{1\ \rm cm^{-3}}\right)^{-1}\frac{I_{\nu}}{1\ \rm Jy\ sr^{-1}},
\end{equation}
to convert the free-free intensity into column density in each pixel, at the frequency $\nu = \rm 353\ GHz$, and the electron temperature of $T_{e} =\rm 8000\ K$. 
This equation shows that the \ion{H}{ii} column density is inverse proportional to the effective density of electrons $n_{\rm e}$, thus we chose $2\ \rm cm^{-3}$ and $10\ \rm cm^{-3}$ \citep{Sodroski97} to calculate the upper and lower limits of the \ion{H}{ii} column density.

\begin{figure*}[ht]
\centering
\includegraphics[scale=0.25]{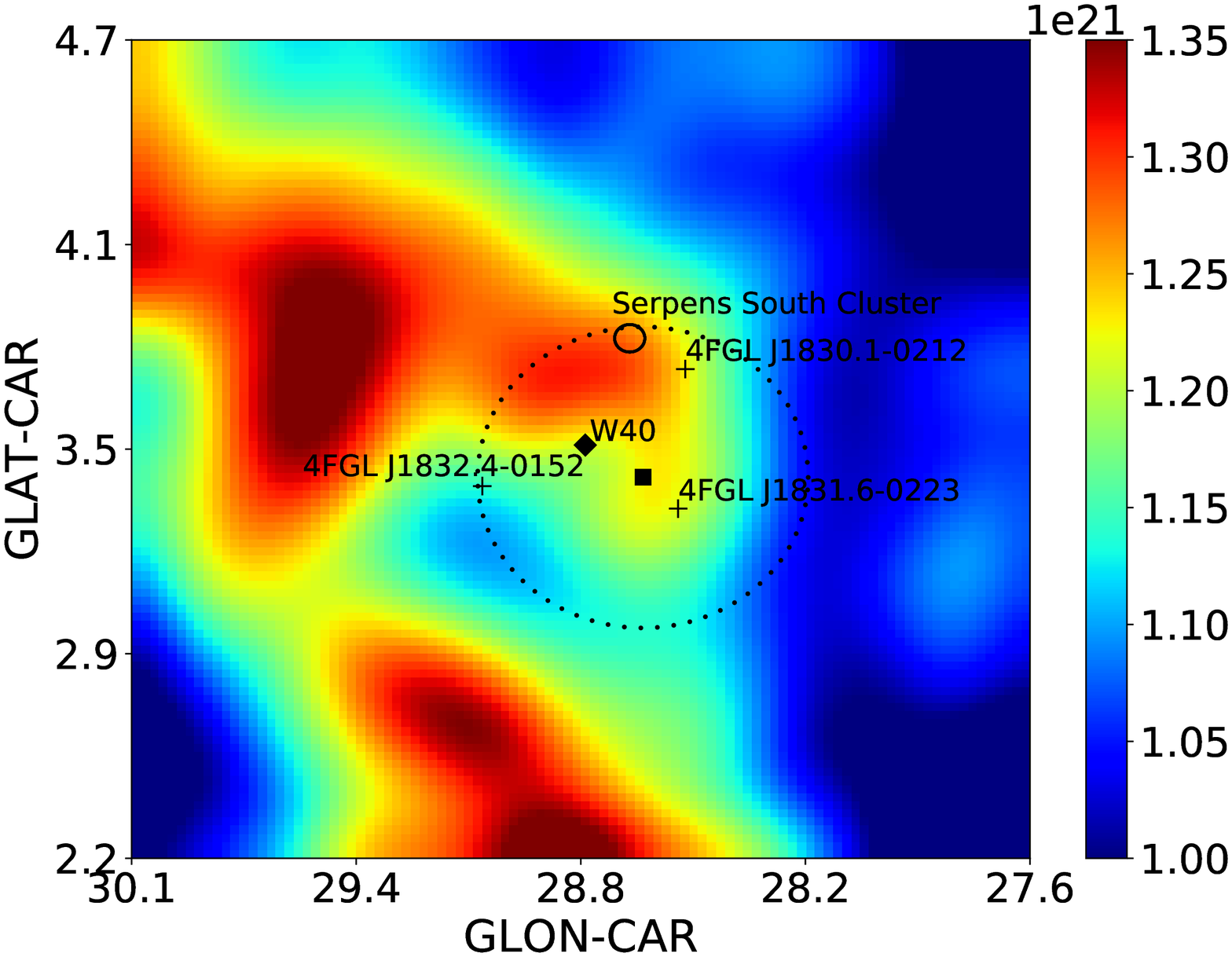}
\includegraphics[scale=0.25]{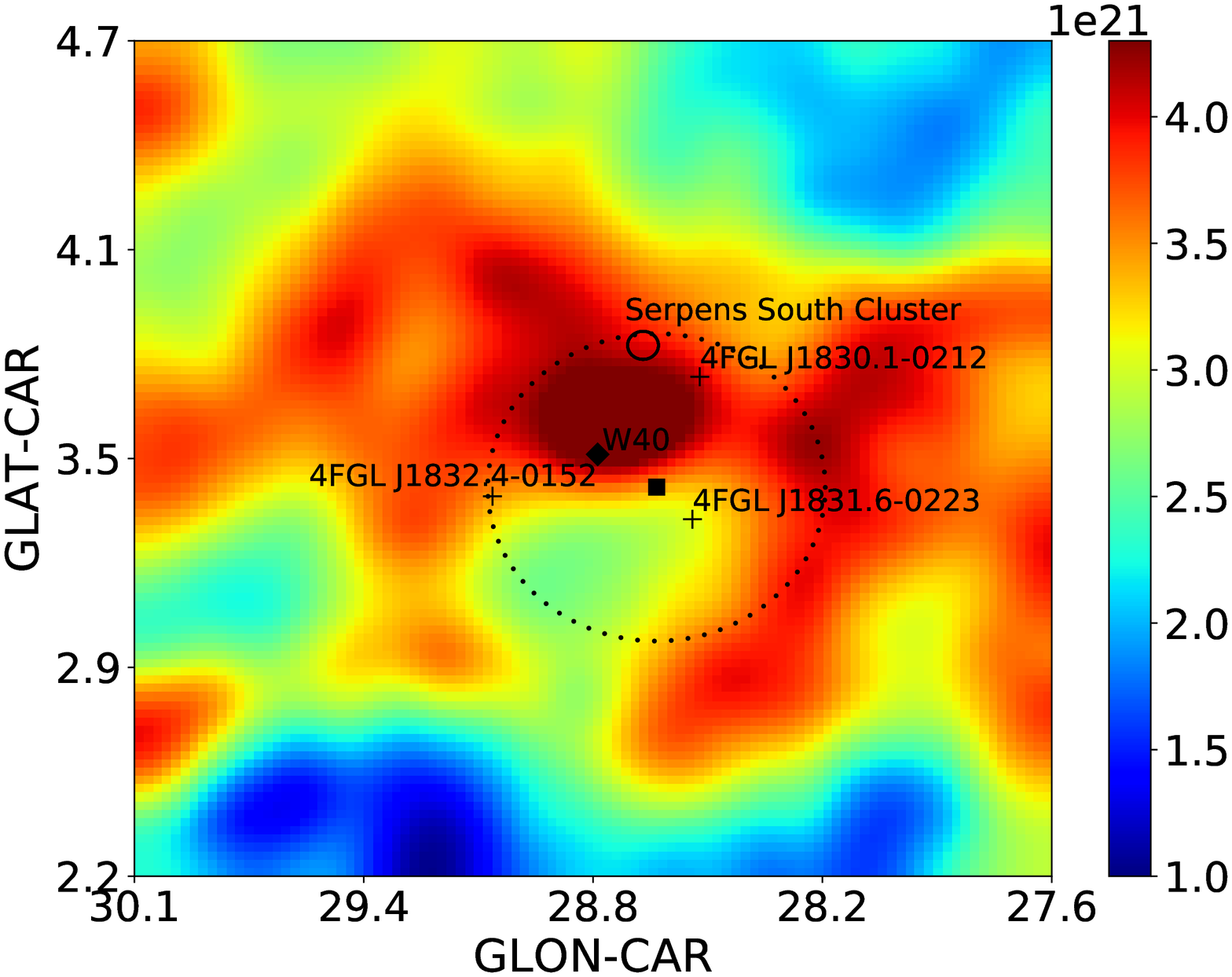}
\includegraphics[scale=0.25]{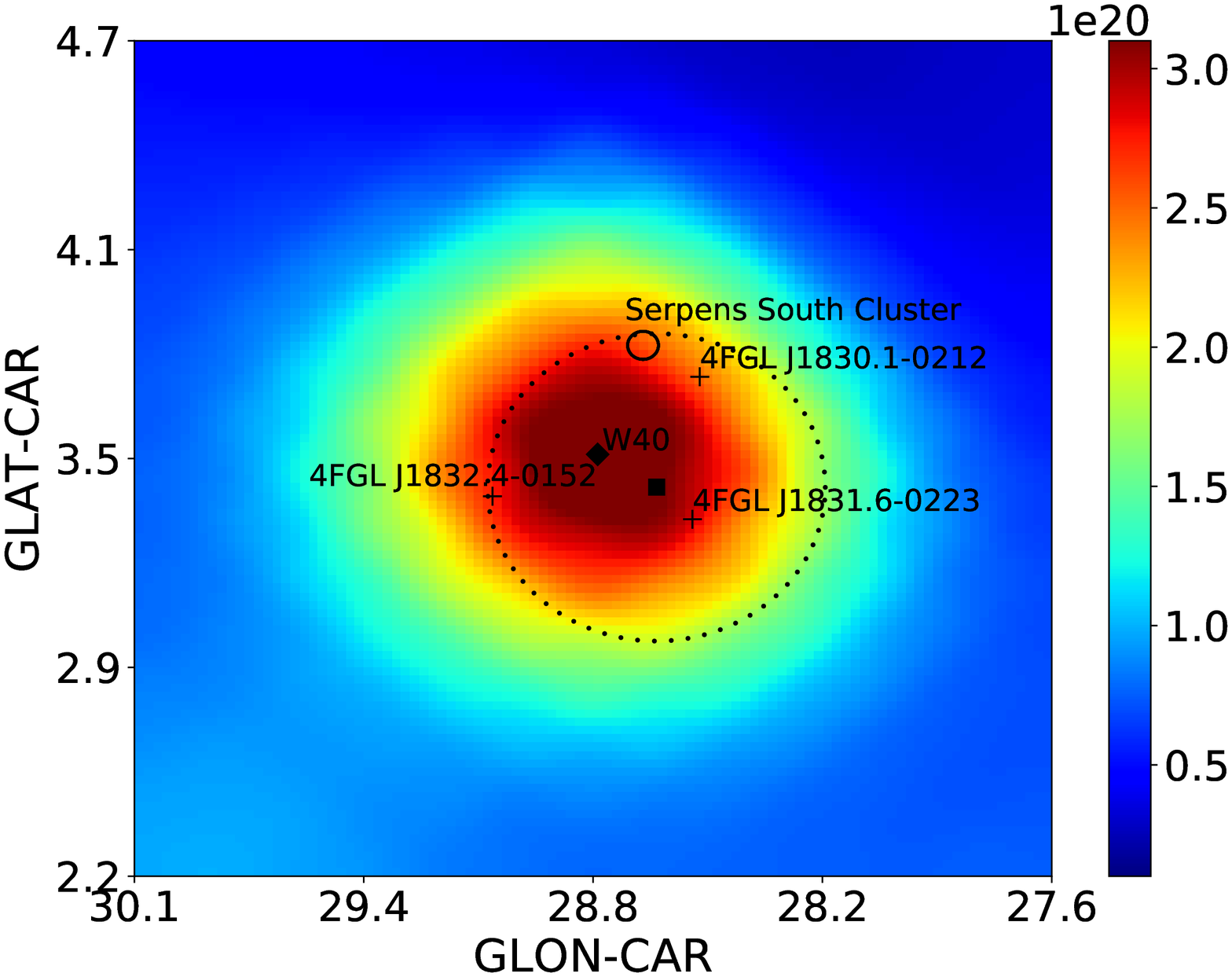}
\caption {
Maps of gas column densities in three phases, smoothed with a Gaussian kernel of $0.6^ \circ$.
The marks are the same as in \figurename~\ref{fig:tsmap}.
Left: the \ion{H}{i} column density derived from 21-cm all-sky survey.
Middle: the H$_{2}$ column density derived from the CO data.
Right: the \ion{H}{ii} column density derived from the Planck free-free map.
For details, see the context in Sect.~\ref{sec:Gas}.
}
\label{fig:gasdis}
\end{figure*}

The derived maps of gas column densities in three phases are shown in \figurename~\ref{fig:gasdis}. We note that, similar to the case of the young massive cluster NGC 3603 \citep{Yang17}, the \gray emission region shows good spatial consistency with the \ion{H}{ii} column densities, which supports the hypothesis that the \gray emission comes from the interactions of the accelerated particles  in the superbubble.

The total mass of the cloud in each pixel can be obtained from the expression
\begin{equation}
M_{\rm H} = m_{\rm H} N_{\rm H} A_{\rm ang} d^{2}
,\end{equation}
where $m_{\rm H}$ is the mass of the hydrogen atom, and $N_{\rm H} = N_{\ion{H}{ii}} + 2N_{\rm H_{2}} + N_{\ion{H}{i}}$ is the total column density of the hydrogen atom in each pixel.  
$A_{\rm ang}$ is the angular area, and $d$ is the distance of the W40 complex.
We calculated the total mass and number of hydrogen atoms in each pixel. The total mass in the \gray emission region is estimated in the range of $1.47 \times 10^{2} \msun < M < 5.49 \times 10^{3} \msun$ as listed in \tablename~\ref{tab2}.
We note that the uncertainty of the gas mass is significant, this is because we cannot exclude the possibility that the \gray emissions are associated only with the \ion{H}{ii} gas due to the compact nature of the \gray emission region (pc scale), and the mass of the \ion{H}{ii} gas is an order of magnitude lower than the total gas mass.
%
Assuming spherical geometry of the \gray emission region, its radius is estimated as $r = d \times \theta \sim 436\ {\rm pc} \times (0.46\deg \times \pi/180\deg)\ {\rm rad} \sim 3.5\ {\rm pc}$. Thus, the average gas number density over the volume should be within $\rm 30\ cm^{-3} < n_{gas} < 1200\ cm^{-3}$.  Because of the good correlation of \gray emission with the \ion{H}{ii} gas, below we used $\rm  n_{gas} \sim 30\ cm^{-3}$  as a fiducial value, but kept in mind that strictly speaking it should be regarded as a lower limit.

\begin{table*}
\centering
\caption{Total mass of the hydrogen atom derived from different tracers within the \gray emission region.}
\begin{tabular}{lccc}
\hline
\hline
Tracer & Gas phase & Mass ($10^{2}\msun$) \\
\hline
ff\tablefootmark{a} ($n_{\rm e} = 2\ cm^{-3}$) & \ion{H}{ii} & 7.35\\ 
ff ($n_{\rm e} = 10\ cm^{-3}$) & \ion{H}{ii} & 1.47 \\ 
21-cm + 2.6-mm line + ff ($n_{\rm e} = 2\ cm^{-3}$) & \ion{H}{i}+H$_{2}$+\ion{H}{ii}  & 54.49 \\ 
21-cm + 2.6-mm line + ff ($n_{\rm e} = 10\ cm^{-3}$) & \ion{H}{i}+H$_{2}$+\ion{H}{ii} & 48.52  \\ 
\hline
\hline
\end{tabular}
\tablefoot{
\tablefoottext{a}{free-free intensity (ff).}
See Sect.~\ref{sec:Gas} for details.
}
\label{tab2}
\end{table*}

\section{The origin of  \gray emission}
\label{sec:cr}
 In principle, it is possible that some \gray residuals  arise because of the imperfect modeling of the Galactic diffuse \gray background. In particular, we note that the \ion{H}{ii} component of gas is not taken into account in the Fermi diffuse background models \citep{fermi_diffuse}. However, as is shown in Fig. \ref{fig:sed}, the predicted  \gray flux for the  \ion{H}{ii} gas (assuming that the CR spectra therein are the same as the local measurement \citep{Aguilar15})  is significantly lower than the observed \gray  flux. Thus, the derived  \gray flux from  this region cannot be caused by uncertainties in the modeling of the diffuse background.  
Two pulsars are located about $0.5^{\circ}$ away from the center of the \gray emission region (as shown in \figurename~\ref{fig:tsmap}), PSR J1830-0131 and PSR J1833-0209 \citep{atnf}. The region PSR J1830-0131 has a distance of 3.5 kpc and a spin-down luminosity of $2.3\times10^{34} ~\rm erg/s$, while for J1833-0209, the values are 13.3 kpc and $4.4\times10^{33} ~\rm erg/s$, respectively. 
If we assume that the \gray emission comes from the pulsar wind nebulae associated with these two pulsars,  the derived \gray luminosities would be about $5\times 10^{34}\rm erg/s$ and $6\times 10^{35}\rm erg/s$, respectively. They significantly exceed the spin-down luminosities of the pulsars. This makes the association of the \gray emission to these pulsars rather unlikely. On the other hand, there are no known SNRs inside this region \citep{green_catalog}. Thus the \gray emission most likely originates from W40 itself. 

We used Naima\footnote{\url{http://naima.readthedocs.org/en/latest/index.html#}} \citep{naima} to fit the SEDs. Naima is a numerical package that allows us to implement different functions and includes tools to perform Markov chain Monte Carlo (MCMC) fitting of nonthermal radiative processes to the data.
In the hadronic scenario, we attribute the observed \grays to the decay of neutral pions produced by the interactions of protons with the ambient gas, using the parameterization of the cross-section  of \citet{Kafexhiu14}. 
Since the low-energy data points are not very constrained, we assume a  power-law distribution in the momentum of the  protons.
The average number density of the target protons is assumed to be $\rm 30\ cm^{-3}$, which is the lower limit derived from the gas distributions above. 
As shown in \figurename~\ref{fig:sed}, the derived index is $\alpha = 2.55 \pm 0.14$, and the total energy is $\rm W_{p} = (2.1\pm{0.5}) \times 10^{46}~\rm erg$ for the protons above 10 GeV. It should be noted that, due to the good spatial correlation of \gray emission and \ion{H}{ii} maps, it is very probable that the \grays are only related to \ion{H}{ii} gas. In this case, as is shown in \figurename~\ref{fig:sed}, the derived CR density is about 10 - 50 times that of the local CRs. 

\begin{figure}[ht]
\centering
\includegraphics[scale=0.4]{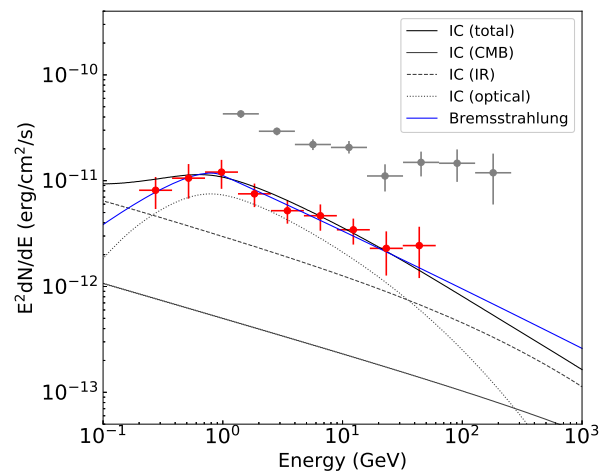}
\caption {
Same as \figurename~\ref{fig:sed} but for leptonic modeling.
}
\label{fig:sed_leptonic}
\end{figure}

We also tested the leptonic scenario assuming the \grays are generated via the relativistic electrons inverse Compton (IC) scattering of low-energy seed photons, or through the nonthermal bremsstrahlung of the relativistic electrons or matter around the W40 region.
For the interstellar radiation field of the IC, we considered the CMB, infrared, and optical emissions calculated by \citet{Popescu17}. The target particle density is assumed to be $\rm 30\ cm^{-3}$ for the relativistic bremsstrahlung. We adopted the formalism described in \citet{Khangulyan14} to calculate the IC spectrum, and for the relativistic bremsstrahlung spectrum we used the parameterization in \citet{Baring99}.  To fit the lower energy break in the \gray spectrum, we should require a relevant break in the spectrum of parent electrons.  Thus, we assumed a broken power law distribution of the relativistic electrons. 
As shown in \figurename~\ref{fig:sed_leptonic}, both the IC and bremsstrahlung model can fit the observable data well, and the corresponding maximum-likelihood values are -4.3 and -1.7.   For IC, the derived parameters for the electrons are $\alpha_{1} = -1.70^{+0.40}_{-0.20}$, $\alpha_{2} = 3.69^{+0.20}_{-0.14}$, $\rm E_{break} = 7.30 \pm 0.90\ GeV$, and total energy $\rm W_{e} = (1.8 \pm 0.2) \times 10^{48}\ erg$. For bremsstrahlung, $\alpha_{1} = -0.20 \pm 0.50$, $\alpha_{2} = 2.65 \pm 0.10$, $\rm E_{break} = 1.01 \pm 0.14\ GeV$, and $\rm W_{e} = (2.1 \pm 0.2) \times 10^{46}\ erg$.  In the regime of bremsstrahlung, the high-energy index of parent electrons $\alpha_{2}$ is similar to the \gray spectral index $\gamma$, while in the IC regime the relation is different, $\gamma = \frac{\alpha_{2}+1}{2}$. 

We cannot formally rule out the leptonic origin of this source. In such a dense region, the bremsstrahlung dominates the radiation mechanism. The good spatial coincidence of the \gray emission with the {\ion{H}{ii}} region  favors a hadronic or bremsstrahlung origin. However,  in the bremsstrahlung scenario, the derived parent proton index changed from -0.2 to 2.65 below and above about 1~GeV. Such a sharp break is quite unusual and  not compatible with any known mechanisms.

If the \gray emission is dominated by the pion decay process, we can use the \gray emissions to study  the CR proton content in this region. To do this, we calculated the \gray emission under the assumption that the CR density around the W40 region coincides with the density in the Solar System measured by AMS-02 \citep{Aguilar15}, with an nuclear enhancement factor of 1.8.
We compared the predictions with the SED derived from the \fermi observations shown in \figurename~\ref{fig:sed}, and the local CR fluxes are lower and the energy spectra are softer than the observed \fermi data. Thus, the CR proton spectrum in this region is significantly harder than the local observations, whilst the CR energy density above 10~GeV is an order of magnitude higher than that of the local CRs when we attribute the \gray emission to the {\ion{H}{ii}} region.

\section{CR content in the vicinity of W40}\label{sec:pro}
If the massive star clusters are indeed CR accelerators, the CRs will inevitably escape and distribute in the vicinity of the massive star clusters. The distribution of CRs can be derived from the \gray emission and gas distributions. In Westerlund 1, the Cygnus Cocoon \citep{Aharonian19}, and Westerlund 2 \citep{Yang18}, the $1/r$ CR profiles are derived, which implies the continuous injection and diffusion dominated propagation of CRs from these massive star clusters. 

\begin{figure*}[ht]
\centering
\includegraphics[scale=0.5]{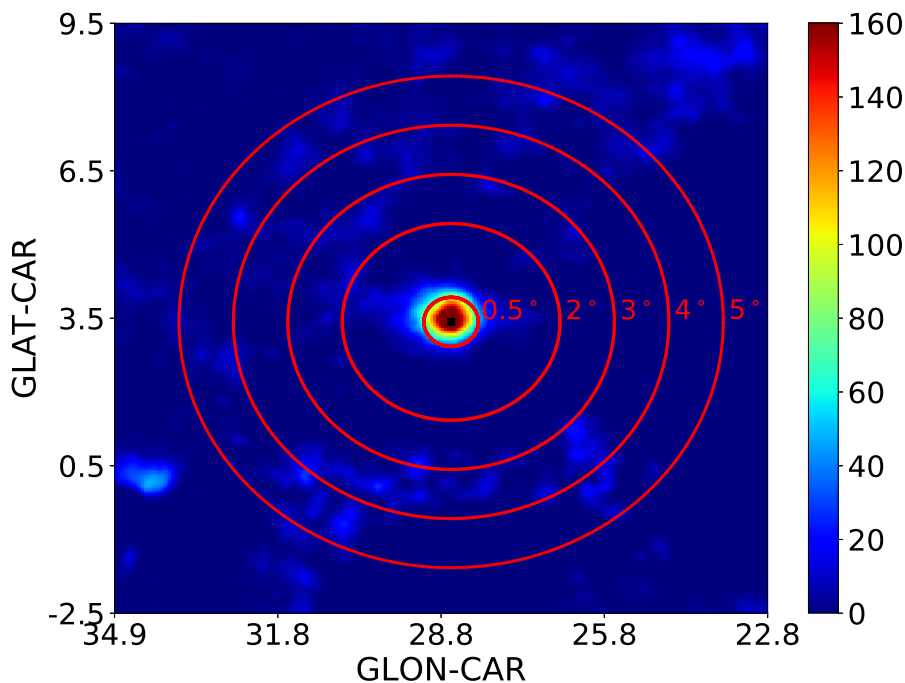}
\caption {TS residual map above 1 GeV overlaid with the rings used to derived upper limits of cosmic ray energy density in Sect.~\ref{sec:pro}.}

\label{fig:ring}
\end{figure*}

\begin{figure*}[ht]
\centering
\includegraphics[scale=0.8]{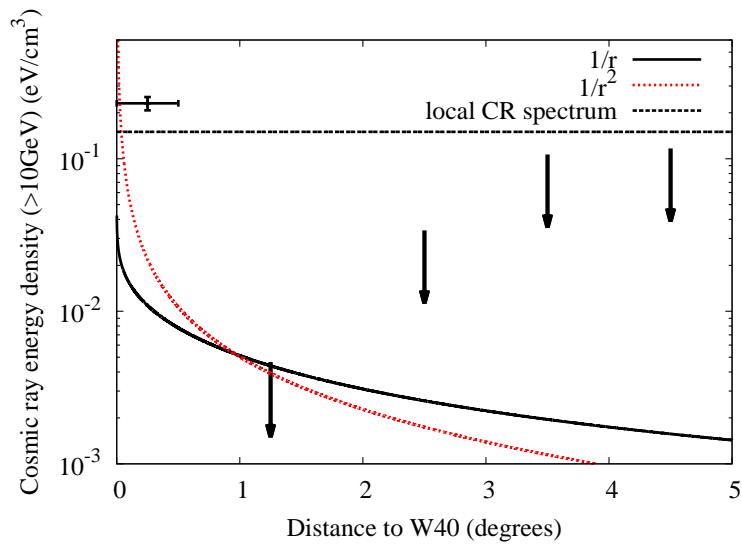}
\caption {Derived CR density profile near W40. The data points are the \gray emission above 1 GeV of W40. The upper limits are derived for the rings defined in Fig.\ref{fig:ring}. The balck and red curves are the projected $1/r$ and $1/r^2$ profiles, respectively. }
\label{fig:pro}
\end{figure*}

For W40, the size of the detected \gray emission region is only of several parsecs, thus probably dominated by the emission from the cluster itself, rather than the cocoons illuminated by the propagating CRs.  To study the propagated CR contents in the vicinity, we divided the inner five-degree region from W40 into four rings, with radii of [0.5:2], [2:3], [3:4], and [4:5] degrees. The TS maps in these regions are shown in Fig.\ref{fig:ring}. Due to the lack of the \gray emission, we derived the 99\% \gray upper limits in each ring. We also derived the corresponding gas mass as in Sect.~\ref{sec:Gas}. Using the \gray production cross-section \citep{Kafexhiu14}, we derived the CR density upper limits profile from the \gray upper limits and gas distributions, the results are shown in Fig.\ref{fig:pro}.  It seems that the second bin gives stringent constrain on the CR density. It can hardly be compatible with the \gray emission detected in W40 itself if we assume a $1/r$ profile. However, it is also possible that the \gray emission in W40 itself can not be regarded as the beginning of the CR profile. Indeed, the CRs produced in W40 may be confined inside the source due to the much slower diffusion in the source region, which forms the \gray emission region. 
The CR luminosity  $L_{\rm p}$ can be estimated  from \gray luminosity $L_{\gamma}$ as 
\begin{equation}
 L_{\rm p}=3 L_{\gamma} (T_{\rm pp}/T_{\rm conf}),
\end{equation}
where $T_{\rm pp}=4\times 10^{13}( \frac{30 ~\rm cm^{-3}}{n} ) \rm s$ is the CR cooling time through pp collision and $T_{\rm conf}=R/2D$ is the confinement time in the \gray emission region. Here,  $n$ is the ambient gas density, $R$ is the size of the \gray emission region, and  $D$ is the diffusion coefficient.  Formally, $D$ can be expressed as $D=k D_{\rm B}=k r_{\rm g} c/3=3\times 10^{22} k (\frac{10~\rm \mu G}{B}) ~\rm cm^2/s $, where $r_{\rm g}$ is the gyro radius, $D_{\rm B}$ is Bohm diffusion coefficient, and $k$ labels the deviation from Bohm diffusion. Thus, the CR luminosity can be expressed as 

\begin{equation}
 L_{\rm p}=3 \times 10^{33} (k/100)(n/30 ~\rm cm^{-3})^{-1}(B/10~\rm \mu G)^{-1} ~\rm erg/s. 
\end{equation}

If $T_{\rm pp}$ is much larger than $T_{\rm conf}$, most of the accelerated CRs will gradually diffuse out the source region. Assuming a continuous injection and the CR density scale as $1/r$, the energy density at the second bin in Fig.\ref{fig:pro} can be estimated as  \citep{aa96}
\begin{equation}
\begin{split}
 w_{\rm p}=L_{\rm p}/4\pi RD=3\times 10^{-4} (k/100)(n/30 ~\rm cm^{-3})^{-1}\\
 (B/10~\rm \mu G)^{-1}(D_{\rm ISM}/10^{28} ~\rm cm^2/s)^{-1} ~\rm eV/cm^3,
 \end{split}
\end{equation}
where $ w_{\rm p}$ is the CR energy density, $D_{\rm ISM}$ is the diffusion coefficient in the interstellar medium, and $R$ is $15$ pc, which is the distance from the second bin to W40. Compared with the derived CR energy density of $5\times10^{-3} ~\rm erg/cm^3$ for the second bin in Fig.\ref{fig:pro}, the derived $k$ should be smaller than about 1000, which can be regarded as a hint that the diffusion in the vicinity of W40 is much smaller than the average value in the interstellar medium (in which $k$ can be as large as $10^6$).  

 The \gray upper limits  reveal a slow diffusion region in the vicinity of W40. This is not surprising, considering the possible  strong turbulence and effective CR confinement surrounding the CR accelerators \citep[see, e.g., ][]{malkov13}. This is also in consistency with our assumption that the detected \gray emissions are from the star cluster itself.





\section{Discussion and conclusion}\label{sec:conc}

Recently,  \citet{Aharonian19}  proposed that young star clusters can be an alternative source population of  Galactic CRs, and the \gray emissions around such objects can be powerful tools to diagnose the acceleration of CRs  and propagation of CRs in the vicinity of sources.  For now, there are already several such systems that have been detected in \gray band, such as  the Cygnus cocoon \citep{Ackermann11, Aharonian19}, Westerlund 1 \citep{Abramowski12}, Westerlund 2 \citep{Yang18}, NGC 3603 \citep{Yang17}, and 30 Dor C \citep{Abramowski15}. Here, we report a statistically significant detection of an extended \gray signal from the direction of another  young star-forming region, W40.  Like the other systems, the spectrum of this source is harder than the local CRs. We argue that the most likely origin of the detected emission is the interactions of CRs accelerated in the young star cluster with the surrounding ionized gas.

Compared with other systems, W40 is unique considering that it is located only 400 pc away from the Solar System and is extremely young at less than 1 Myr. Within such a young star forming system, no stars can have enough time to evolve into a supernova. Thus, if the \grays are illuminated by CRs accelerated in W40, the only feasible acceleration sites for CRs are the stellar winds of young massive stars. Furthermore, it is  the first time the \gray emission has been detected in pc scale, rather than the 100 pc cocoons observed in other systems. Additionally, the size of the detected \gray emissions is not far from the stellar cluster itself. Thus, it is  likely that the observed \grays are directly from the stellar cluster.   

The total CR energy in W40 is only to the order of $\rm 10^{47}~erg$, which is much lower than that derived from the similar systems such as NGC 3603 \citep{Yang17}, Westerlund 2 \citep{Yang18}, and the Cygnus cocoon \citep{Ackermann11, Aharonian19}.  However, the determination of the total CR  energy from \gray emissions is not decisive. It is possible that  CRs  are distributed in a much larger area with a much lower gas density. In this case, if the CRs are distributed as $1/r$ as derived  from other stellar clusters \citep{Aharonian19}, the total CR energy scales as $R^2$, where $R$ is the maximum distance penetrated by a particle accelerated by W40. Thus, the total CR energy can be orders of magnitude larger if $R$ is dozens of parsecs, as in other systems. Also, W40 is much less powerful than the other detected systems. There are only four identified OB stars, compared with more than a dozen in other systems. Therefore, it is reasonable to assume the wind power in W40 can be orders of magnitude lower. In this case, it is not surprising that W40 reveals a much lower CR luminosity. Meanwhile, the derived index of CR proton spectrum ($\sim 2.5$) is also slightly softer than that in other observed massive star clusters ($\sim 2.2 - 2.3$). This may be also due to the younger age of this star cluster, as well as a lower wind power compared with other systems that were observed at greater distances. Another possibility is that the difference reflects  the different CR spectra in the star cluster and in the surrounding cocoon, which can be induced by the propagation effects. A dedicated study of the mechanisms of acceleration and propagation, as well as a careful multiwavelength observations to estimate the total wind power in such systems are required to explain such differences. Also, if there are other objects like W40 but located at greater distances, they may be difficult to be resolved by current \gray observations. In this case, however, such objects may contribute significantly to the diffuse \gray emission. Thus, further caution is required to extract the CR information from diffuse \gray emissions \citep[e.g., ][]{fermi_diffuse, yang16}.

\section*{Acknowledgements}
This work is supported by  the NSFC under grants 11421303,11625312 and 11851304 and the National Key R \& D program of China under the grant 2018YFA0404203. Ruizhi Yang  is supported  by the national youth thousand talents program in China.

\bibliographystyle{aa}
\bibliography{ms}

\end{document}